\documentclass{article}
\usepackage{graphicx}
\usepackage{subcaption}
\usepackage{authblk}
\begin{document}

\bibliographystyle{plain}

\title{Testing Linearly Coasting Cosmology by Strong Lensing System and Pantheon+ Data}
\author{Savita Gahlaut}
\affil{\textit{Deen Dayal Upadhayaya College, University of Delhi} \\ \textit{Sector-3, Dwarka, New Delhi 110078, India} \\ \textit{E-mail : savitagahlaut@ddu.du.ac.in}}
\date{}
\maketitle
\begin{abstract}
The standard model of cosmology ($\Lambda$CDM) is facing a serious crisis caused by the inconsistencies in the measurements of some fundamental cosmological parameters (Hubble constant $H_{0}$ and cosmic curvature parameter $\Omega_{k}$ for example). On the other hand, a strictly linear evolution of the cosmological scale factor is found to be an excellent fit to a host of observations. Any model that can support such a coasting presents itself as a falsifiable model as far as the cosmological tests are concerned. 
In this article the observational data of strong gravitational lensing (SGL) systems from SLACS, BELLS, LSD and SL2S surveys has been used to test the viability of linearly coasting cosmology. 
Assuming the spherically symmetric mass distribution in lensing galaxies, the ratio of angular diameter distance from lens to source and angular diameter distance of the source is evaluated and is used to constrain the power law cosmology. Further, updated  type Ia supernovae dataset (Pantheon+) with covariance matrix incorporating all statistical and systematic uncertainties is used to constrain the power law cosmology.  
It is found that the linear coasting is consistent with the SGL data within 1-$\sigma$ uncertainties but  Pantheon+ sample does not support linear coasting. 
\end{abstract}

\newpage
\section{Introduction}
At present there is a general consensus that the $\Lambda$CDM model of cosmology \cite{Peebles1984} is most consistent with the cosmological observations and is considered to be the standard model of cosmology. The model describes a spatially flat universe predominantly filled with dark energy in the form of a cosmological constant $\Lambda$ and cold dark matter. These two comprise about $95\%$ of the total energy of the universe. Fundamental cosmological parameters in the standard model are constrained using a host of observational data. Recently different independent data sets used to constrain some fundamental cosmological parameters, Hubble constant $H_{0}$ (a parameter that describes the present expansion rate of the universe) and cosmic curvature parameter $\Omega_{k}$ (a parameter that determines the geometry of the universe) for example, shows remarkable inconsistencies in the estimates of these parameters. Measurements of cosmic microwave background radiation (CMBR) temperature and anisotropies from \textit{Planck} satellite estimates a value of $H_{0} = 67.4 \pm 0.5$ $km$ $s^{-1}$ $Mpc^{-1}$ in  $\Lambda$CDM model \cite{Aghanim}. Whereas, type Ia supernovae (SNe Ia) probes predict    
 $H_{0} = 74.03 \pm 1.42$ $km$ $s^{-1}$ $Mpc^{-1}$ by Hubble-Lema$\hat{i}$tre law without assuming any cosmological model \cite{Riess}. The $4.4\sigma$ tension between the two independent estimates can not be attributed to systematic error and revels the inconsistency between the early universe and the late universe   \cite{Riess,Vale,Feeney,Verde}. 
 Similarly, the estimates of the cosmic curvature parameter $\Omega_{k}$ from CMBR temperature and polarization measurements from \textit{Planck} suggest a closed universe ($\Omega_{k} < 0$) at $99\%$ confidence level \cite{Aghanim,Vale19} whereas the combination of CMBR data and baryon acoustic oscillation (BAO) measurements predicts a spatially flat universe ($\Omega_{k}=0$) with a remarkable $0.2\%$ precision \cite{Handley}. These inconsistencies indicate that either some new physical phenomenon are at play which are not well understood or the standard model is flawed and other cosmological models of the universe are worth exploring.

 Power law cosmology with the scale factor $a(t) \propto t^{\beta}$, where $\beta$ is a constant, is found to be an excellent fit to a host of observations with $\beta \approx 1$. Linearly coasting cosmology (Power law cosmology with $\beta \approx 1$) is found to be comfortably concordant with various low redshift probes such as SN-Ia \cite{SN1,SN2,SN3}, quasar angular sizes (QSO) \cite{qso}, cosmic chronometer H(z) \cite{hz}, gravitational lensing statistics \cite{lensing}, BAO \cite{bao,bao1}, galaxy cluster gas mass fractions \cite{massfraction} and the combination of H(z)+ BAO + SN-Ia + gamma-ray burst distance moduli (GRB) \cite{GRB}. Such a model of the universe is free from the horizon problem and flatness problem. The age of the universe in such a model is consistent with the age estimates for old stars and globular clusters \cite{SN1,hz,age}. Further, nucleosynthesis in a linearly coasting universe produces the desired amount of primordial Helium along with the metallicity observed in the lowest metallicity objects \cite{Nucleosynthesis,nuc}. Further, linear evolution of the scale factor is supported in alternative non-minimally coupled gravity theories where it turns out to be independent of the equation of state of matter \cite{nonmin,nonmin1,nonmin2}. The coupling of the large scale scalar curvature of the universe to the scalar field in non-minimally coupled theories give rise to a characteristic evolution: the non-minimal coupling diverges, the scale factor approaches linearity and the non-minimally coupled field acquires a stress energy that cancels the vacuum energy in the theory. This aspect has been widely explored in attempts to solve the cosmological constant problem. 
 
Linearly coasting cosmology presents a simpler alternative to the expansion history of the universe predicted by the standard model which can be falsified if it fails to provide a good fit to the available observational data. In recent years strong gravitational lensing has become a very important astrophysical tool to estimate cosmological parameters. Earlier attempts to use strong lensing to constrain cosmological parameters were based on two approaches.
First was the statistical one, using CLASS or SQLS samples \cite{chae,oguri} , based on comparison between empirical distribution of image separations in observed samples of lenses and the theoretical one. The other method used galaxy clusters as lenses \cite{pac,sereno,gilmore,jullo} with each lens generating multiple images of various sources i.e. background galaxies. With better understanding of structure and evolution of early type galaxies to make assessment of mass density profile and availability of reasonable catalogs of strong lenses with spectroscopic and astrometric data (obtained with well-defined selection criteria), the ratio of angular diameter distance from source to lens and from source to observer is estimated and used to constraint the cosmological parameters\cite{bies,cao}.
In this paper, I test the viability of linearly coasting cosmology by constraining the power law cosmology using the strong gravitational lensing data of 118 lenses from SLACS, BELLS, LSD and SL2S surveys compiled by Cao et al.\cite{Cao}. 

Type Ia supernovae (SNe Ia) are considered as the standard candle in cosmology due to their superior brightness, at their peak matching the brightness of a typical galaxy. These are often adopted to provide the most effective method to measure cosmological distances.
 Recently, the largest combined SNe IA sample referred to as Pantheon+ data set is released by Scolnic et al. \cite{Scol}. The sample consists of 1701 data points in the redshift range $0.001 <z< 2.3$. I analyze the Pantheon+ data set to obtain constraints on power law cosmology.

The paper is organized as follows. In Section 2, I describe the power law cosmology ansatz and derive the angular diameter distance and luminosity distance relations.
In Section 3, I briefly describe the method to estimate angular diameter distance using strong gravitational lensing and how SGL data set and SNe Ia dataset is used to constrain the power index for the model. In section 4 methodology and data is discussed. Finally, the results are summarized in section 5.
   
\section{Power Law Cosmology}
Large scale homogeneity and isotropy observed in the universe suggests the geometry of the universe can be described by the Friedmann-Lema$\hat{i}$tre-Robertson-Walker (FLRW) metric:
\begin{equation}
ds^{2} = c^{2}dt^{2}-a(t)^{2}[\frac{dr^{2}}{1-K r^{2}} + r^{2}(d\theta^{2}+ \sin^{2}{\theta} d\phi^{2})]
\end{equation}
 Here $c$ is the speed of light, $t$ is the proper time, $r$, $\theta$ and $\phi$ are the spherical polar co-moving coordinates. $K$ is the curvature constant and is $\pm 1$ or  $0$ for a suitable choice of units for $r$. $a(t)$ is an unknown function of $t$ and is called the cosmic scale factor or expansion parameter. In standard model the scale factor and curvature constant specify the dynamics of the universe and are determined from the Einstein's equation from the general theory of relativity (GTR) for a homogeneous and isotropic fluid as source.
 
 In power law cosmology, the scale factor $a(t)$ takes the form:
\begin{equation}
a(t) = \alpha t^{\beta}
\end{equation} 
where $\alpha$ and $\beta$ are constants. The Hubble parameter (defined as $H(t) = \frac{\dot{a}}{a}$) is:
\begin{equation}
H(t) = \frac{\beta}{t}
\end{equation}
From the definition of redshift, $\frac{a_{0}}{a(t)} \equiv 1+ z$, where $a_{0}$ is the current value of scale factor and $z$ is the redshift, we have
$$a(t) = \alpha t^{\beta} = \frac{a_{0}}{1+z}$$
$$\Rightarrow \frac{1}{t} = [\frac{\alpha}{a_{0}}(1+z)]^{\frac{1}{\beta}}$$ and
\begin{equation}
H(z) = H_{0}(1+z)^{\frac{1}{\beta}}
\end{equation}
where $H_{0} = \beta(\frac{\alpha}{a_{0}})^{\frac{1}{\beta}}$ is the present value of the Hubble constant.

The dimensionless co-moving distance $d(z)$ in FLRW cosmology is:
\begin{equation}
d(z) = \left\lbrace\begin{array}{cc} D_{c},&K=0\\ &\\
\frac{a_{0}H_{0}}{c}\sinh{(\frac{c}{a_{0}H_{0}}D_{c})},&K=-1\\ &\\ \frac{a_{0}H_{0}}{c}\sin{(\frac{c}{a_{0}H_{0}}D_{c})},&K=1

\end{array}\right.
\end{equation}
where $D_{c} = \int_{0}^{z}\frac{H_{0}}{H(z')}dz'$. The angular diameter distance is 
\begin{equation}
D_{A}(z) = \frac{c}{H_{0}(1+z)} d(z)
\end{equation}
and the luminosity distance is
\begin{equation}
D_{L}(z) = (1+z) \frac{c}{H_{0}} d(z)
\end{equation}

\section{Strong Gravitational Lensing}
Strong gravitational lensing (SGL) occurs when the multiple images of a background galaxy (source) appear due to lensing effect of a galaxy or cluster of galaxies (lens) lying along its line of sight. The multiple image separation of the source in a specific strong lensing system depends only on angular diameter distances to the lens and to the source provided a reliable model for the mass distribution within the lens is known.
A perfect alignment of the source, lens and observer along the same line of sight results in symmetry around the lens and causing a ring like structure called Einstein ring.
The Einstein ring radius $\theta_{E}$ depends on the mass of the lensing object $M_{lens}$; more massive it is, the larger the radius. It also depends on the distance between observer to lens $D_{l}$, observer to source $D_{s}$ and  source to lens $D_{ls}$ and is given by \cite{Sch}:
\begin{equation}
\theta_{E} = \left[\frac{4GM_{lens}D_{ls}}{c^{2}D_{l}D_{s}}\right]^{1/2}
\end{equation}
From stellar kinematic measurements, the dynamical mass $M_{E}$ inside the Einstein radius $R_{E} \equiv \theta_{E} D_{l}$ for lenses having a singular isothermal sphere (SIS) mass distribution is:
\begin{equation}
M_{E} = \frac{\pi }{G}\sigma_{SIS}^{2}R_{E}
\end{equation}
where $\sigma_{SIS}$ is the stellar velocity dispersion of the lens mass distribution.
For lenses having a singular isothermal sphere (SIS) mass distribution, the Einstein radius is:
\begin{equation}
\theta_{E} = 4\pi \frac{\sigma_{SIS}^{2}}{c^{2}} \frac{D_{ls}}{D_{s}}
\end{equation}
 Thus $\theta_{E}$ depends on the cosmological model through the ratio $D_{ls}/D_{s}$, the angular diameter distance between source and lens and between lens and observer. The cosmological model of the universe can be tested using equation (10)
if one has a reliable data of the lensing system, i.e., Einstein radius $\theta_{E}$ from image astronomy and $\sigma_{SIS}$ from central velocity dispersion obtained from spectroscopy \cite{bies,cao}.
With new and powerful upcoming sky surveys along with the ongoing surveys with better precision, SGL is going to be one of the most effective probes to constraint cosmological parameters in a model independent way.

\section{Methodology and Data}
\subsection{SGL Data}
To constraint a cosmological model using SGL data, the quantity to be estimated is:
\begin{equation}
D_{th}(z_{l},z_{s}) = \frac{D_{ls}(z_{l},z_{s})}{D_{s}(z_{s})}
\end{equation} 
and corresponding observable is
\begin{equation}
D_{ob} = \frac{c^{2}}{4\pi}\frac{\theta_{E}}{\sigma_{0}^{2}}
\end{equation}

As the dependence on cosmological model is through a distance ratio, the method is independent of the Hubble constant value $H_{0}$ and is not affected by dust absorption or source evolutionary effects. However it depends upon the reliability of lens modeling and measurements of $\sigma_{0}$. 
In a flat power law cosmology ($K = 0$), the ratio $D_{th}(z_{l},z_{s})$ is:
\begin{equation}
D_{th}(z_{l},z_{s}) = \frac{[(1+z_{s})^{1-\frac{1}{\beta}}-(1+z_{l})^{1-\frac{1}{\beta}}]}{[(1+z_{s})^{1-\frac{1}{\beta}}-1]}
\end{equation} 
and in the limiting case $\beta \rightarrow 1$, 
\begin{equation}
D_{th}(z_{l},z_{s}) = \frac{\log(\frac{1+z_{s}}{1+z_{l}})}{\log(1+z_{s})}
\end{equation}
The corresponding quantities in open power law cosmology (with $K=-1$) are
\begin{equation}
D_{th}(z_{l},z_{s}) =\frac{\sinh[k\frac{\beta}{\beta-1}[(1+z_{s})^{1-\frac{1}{\beta}}-(1+z_{l})^{1-\frac{1}{\beta}}]]} {\sinh[k\frac{\beta}{\beta-1}[(1+z_{s})^{1-\frac{1}{\beta}}-1)]]}
\end{equation} 
where $k = \frac{c}{a_{0}H_{0}}$ and in the limiting case $\beta \rightarrow 1$, the ratio is:
\begin{equation}
D_{th}(z_{l},z_{s}) = \frac{(\frac{1+z_{s}}{1+z_{l}})^{k}-(\frac{1+z_{l}}{1+z_{s}})^{k}}{(1+z_{s})^{k}-(1+z_{s})^{-k}}
\end{equation}

The data sets used in this paper consists of 118 strong lensing systems from Solan Lens ACS Survey (SLAC), BOSS Emission-Line Lens Survey (BELLS), Lenses Structure and Dynamics Survey (LSD)and Strong Lensing Legacy Survey (SL2S). The dataset was compiled by Cao et al.(2015)[Table 1 in \cite{Cao}]. For each lens, the source redshift ($z_{s}$), lens redshift ($z_{l}$) and luminosity averaged central velocity dispersion $ \sigma_{0}$ are determined using spectroscopy from Solan Digital Sky Survey (SDSS), Baryon Oscillation Spectroscopic Survey (BOSS), LSD and Canada France Hawaii Telescope Legacy Survey (CFHTLS). Most of the lensing galaxies in the dataset used are early type elliptical galaxies and a SIS mass profile for such galaxies is strongly supported in various independent studies \cite{Fuk,Hel,Laga, Ruff}.

As the velocity dispersion for a SIS lens $\sigma_{SIS}$ may not be same as the central velocity dispersion $\sigma_{0}$ so a new parameter $f_{E}$ was introduced by Kochanek (1992)\cite{Koch} such that $\sigma_{SIS} = f_{E}\sigma_{0}$. The parameter $f_{E}$ compensates for the contribution of dark matter halos in velocity dispersion as well as systematic errors in measurement of image separation and any possible effect of background matter over lensing systems. All these factors can affect the image separation by up to $20\%$ which limits $\sqrt{0.8} < f_{E} < \sqrt{1.2}$
\cite{cao,Ofek}. In this work $f_{E}$ is taken as a free parameter fitted together with the cosmological parameters $\beta$ and $k$.
The Einstein radius $\theta_{E}$ is determined from Hubble Space Telescope Images (HST). In different surveys considered here, the method to determine Einstein radius is more or less consistent. The Einstein radii were measured by fitting the mass distribution models, after subtracting de Vaucouleurs profile of the lens, to generate lensed image. The fractional uncertainty of $\theta_{E}$ is estimated to be $5\%$ across all surveys \cite{Cao}.

\subsubsection{SNe Ia Data}

The distance modulus $\mu$ of a supernova is defined as:
\begin{equation}
\mu^{th} = m - M' = 5 log(D_{L}/Mpc)+ 25
\end{equation}
where $m$ is the apparent magnitude, $M'$ is the absolute magnitude and $D_{L}$ is the luminosity distance. For the Pantheon+ SNe Ia sample used in this paper, the observed distance modulus $\mu^{ob}$ is determined by using the Spectral Adaptive Light Curve Template 2 (SALT2) and is given by:
\begin{equation}
\mu^{ob} = m_{B} - M_{B} + \alpha X_{1} - \eta \varsigma +\bigtriangleup_{M} + \bigtriangleup_{B}
\end{equation} 
where $m_{B}$  and $M_{B}$ are the apparent magnitude and absolute magnitude of B-band, $X_{1}$ and $\varsigma$ are the curve shape and colour parameters, $\alpha$ and $\eta$ are nuisance parameters of the luminosity-stretch and luminosity-colour relations, $\bigtriangleup_{M}$ and $\bigtriangleup_{B}$ are the distance corrections and bias corrections, respectively. In general, the nuisance parameters $\alpha$ and $\eta$ are fitted along with cosmological parameters. Recently, Scolnic et al. \cite{Scol} analyzed 1701 light curves of 1550 spectroscopically confirmed Type Ia supernovae called Pantheon+ sample. Scolnic et al. reported corrected apparent magnitude $m^{corr} = \mu^{ob} + M_{B}$ for all the SNe in Pantheon+ sample using BEAMS with bias corrections (BBS) method for fitting the nuisance parameters $\alpha$ and $\eta$ \cite{KessSco17}. The uncertainties, including statistical and systematic uncertainties, for the data are given by a $1701$x$1701$ matrix $C$ \cite{Scol}.

Constraints on power law cosmology are obtained by performing a Markov Chain Monte Carlo (MCMC) analysis using Python module \textit{emcee} \cite{emcee} by maximizing the likelihood function \textit{L} $= e^{-\chi^{2}/2}$, to determine the best-fit values of the parameters of the model. 

The likelihood function for the SGL dataset used is evaluated using the expression:
\begin{equation}
\chi_{1}^{2} = \sum_{i=1}^{N} \left(\frac{D_{th}(z_{l,i},z_{s,i},\textbf{p})-D_{ob}(\sigma_{0,i},\theta_{E,i})}{D_{ob}\Delta D_{ob,i}}\right)^{2}
\end{equation}
here $\textbf{p}$ represents cosmological model parameters, $N$ is the total number of data points and $\Delta D_{ob}$ is the uncertainty in the value of $D_{ob}$ given by:
\begin{equation}
\Delta D_{ob} = \sqrt{(\frac{\Delta\theta_{E}}{\theta_{E}})^{2}+ 4(\frac{\Delta\sigma_{0}}{\sigma_{0}})^{2})}
\end{equation}
where $\Delta\theta_{E}$ and $\Delta\sigma_{0}$ are the uncertainties  in the measurement of Einstein radius and velocity dispersion respectively.

For the supernovae Pantheon+ sample the chi-square function is:
\begin{equation}
\chi_{2}^{2} =  \mathbf{\bigtriangleup\mu}^{\dagger} \cdot C^{-1} \cdot \mathbf{\bigtriangleup\mu}
\end{equation}
where $\bigtriangleup\mu_{i} = \mu^{th}(z_{i},\textbf{p}) -  \mu^{ob}(z_{i})$ is the vector of residuals of the sample.

For the joint analysis of both SGL and supernovae samples, the log-likelihood function becomes:
\begin{equation}
ln(L_{tot}) = -0.5(\chi_{1}^{2} + \chi_{2}^{2})
\end{equation}

\section{Results and Discussion}

 The constraints obtained in this work are summarized in Table 1 and Table 2. The best fit values and 1-dimensional marginalized best fit values and uncertainties of parameters were obtained for a flat ($K = 0$) and open ($K = -1$) power law universe.
 The parameters fitted for the flat universe are : $\beta$, $f_{E}$ and $M$, where $M = M_{B} +5 log_{10}(c/ H_{0}Mpc)+25$. Given the degeneracy of factor $5 log_{10}(c/ H_{0}Mpc)+25$ with absolute magnitude $M_{B}$, both are combined and fitted as the parameter $M$. 
  The priors used for all parameters are flat with $\beta$ prior non-zero over the range $0.1\leq \beta \leq 4$, $f_{E}$ prior non-zero over the range $0.8\leq f_{E} \leq 1.2$ and $M$ prior non-zero over the range $ 20 < M< 30$. For the open ($K=-1$)power law cosmology, the datasets do not constrain the parameter $k= \frac{c}{a_{0}H_{0}}$ and it is found that for $k<<1$ the constraints on $\beta$, $f_{E}$ and $M$ are same as that for  the flat case. The same conclusion can also be drawn from equation (15), which shows that for $k=\frac{c}{a_{0}H_{0}} <<1$ the ratio $D_{th}(z_{l},z_{s})$ for the open case is same as that for the flat case. 1-D marginalized best fit values of $\beta$, $f_{E}$ and $M$ are then obtained with non-zero flat prior for $k$ over the range $0.01\leq k \leq 0.5$. The posterior 1-dimensional probability distributions and two dimensional confidence  regions of the parameters for the flat and open power law models are presented in Figs. 1-6.

 The aim of this paper was to test the viability of a linearly coasting cosmology with $a(t) \sim t$, which presents itself as a falsifiable model. The constraints obtained on the general power law cosmology using SGL data shows that a linear coasting is accommodated well within $1\sigma$. However, the best fitting value of $\beta$ from Pantheon+ SNe Ia sample is not consistent with the linearly coasting universe. Joint analysis of SGL + Pantheon+ sample also do not accommodate the linearly coasting universe.
 
  Further, the results of various independent studies using different datasets to constrain  power law exponent $\beta$ are summarized in Table 3.

From the analysis of this work and the results presented in Table 3,
 I conclude that, although linearly coasting cosmology presents a simple evolution of the universe with several good features, at present this class of power law cosmologies is not a flawless model of the universe and can not be considered a viable alternative to the standard $\Lambda CDM$ model.

\begin{table}
\begin{center}

\begin{tabular}{|c|c|c|c|c|c|}  \hline\hline\
K & Data & $\beta$ & $f_{E}$ & k & M \\ \hline

0 & SGL & 0.8443 & 1.0160 & ..&.. \\ 
0 & SNe Ia & 1.3267 & .. &..& 23.8372  \\

0 & SGL+SNe Ia & 1.3197 & 0.9905 &.. & 23.8380 \\
\hline & & & & & \\
-1 & SGL & 0.9489 & 1.0103 & 0.3 & .. \\
-1 & SNe Ia & 1.3267 &..& 0.01 & 23.8372 \\

-1 & SGL+SNe Ia & 1.2523 & 0.9990 & 0.4424 & 23.8414  \\

 \hline \hline
\end{tabular}

\caption{\small Unmarginalized Best fit parameter values.}
\end{center}
\end{table}

\begin{table}

\begin{tabular}{|c|c|c|c|c|c|}  \hline\hline\
K & Data &$\beta$ & $f_{E}$ & k & M \\ 
\hline & & & & & \\
0& SGL & $1.0046^{+0.5465}_{-0.2311}$ & $1.0043^{+0.0199}_{-0.0213}$ &..&.. \\ 
&&&&&\\
0 & SNe Ia & $1.3266^{+0.0508}_{-0.0457}$ & .. & ..& $23.8369^{+0.0065}_{-0.0065}$\\
&&&&&\\
0 & SGL+SNe Ia  & $1.3265^{+0.0504}_{-0.0500}$ &$0.9909^{+0.0073}_{-0.0075}$ & .. &$23.8377^{+0.0064}_{-0.0070}$ \\
&&&&&\\
\hline & & & & & \\
-1 & SGL   & $1.1567^{+0.9607}_{-0.3283}$ & $0.9976^{+0.0217}_{-0.0212}$ &$0.2071^{+0.0701}_{-0.1223}$&..\\
&&&&&\\
-1 & SNe Ia  & $1.3124^{+0.0558}_{-0.0518}$ &.. & $ 0.1482^{+0.1495}_{-0.0945}$& $23.8378^{+0.0068}_{-0.0069}$\\
&&&&&\\

-1 & SGL+SNe Ia  & $1.2881^{+0.0563}_{-0.0512}$ &$0.9954^{+0.0081}_{-0.0085}$ & $0.2819^{+0.1526}_{-0.1889} $ & $23.8392^{+0.0069}_{-0.0066}$ \\

&&&&&\\
 \hline \hline
\end{tabular}

\caption{\small 1-dimensional marginalized best fit parameter values and uncertainties.}

\end{table}

\begin{table}

\begin{center}

\begin{tabular}{|c|c|c|}  \hline\hline\
 Data & $\beta$ & Ref. \\ 
\hline & &  \\

SNe Ia & $1.04^{+0.07}_{-0.06}$ &\cite{SN1} \\
SNe Ia & $1.0042 \pm 0.043$ & \cite{SN2} \\

H(z) & $1.05^{+0.071}_{-0.066}$ & \cite{SN3} \\

SNe Ia & $1.44^{+0.26}_{-0.18}$ & \\
QSO & $1.0\pm 0.3$ & \cite{qso} \\

H(z) & $1.07^{+0.11}_{-0.08}$ &\cite{hz} \\ 
SNe Ia & $1.42^{+0.08}_{-0.047}$ & \\

GL Stat.  & $1.13^{+0.4}_{-0.3}$ &\cite{lensing} \\

SNe Ia & $1.55 \pm 0.13$ & \\
BAO & $1.3$ & \\

BAO & $0.93$ & \cite{bao}\\
SNe Ia & $1.44 - 1.56$ & \\
SNe Ia & $1.55 \pm 0.13$ & \cite{bao1} \\
BAO & $0.908\pm 0.019$ & \\
H(z)+BAO+SNe Ia+GRB & $1.08\pm 0.04$ & \cite{GRB} \\
 
H(z) & $1.013^{+0.06983}_{-0.1038}$ & \cite{Ryan}  \\
BAO & $0.9211^{+0.01653}_{-0.01652}$ & \\
QSO & $1.045^{+0.1142}_{-0.2054}$ & \\
GRB & $0.8707^{+0.1197}_{-0.2782}$ & \\
HIIG & $1.310^{+0.1219}_{-0.1988}$ & \\

SNe Ia & $1.52 \pm 0.15$ & \cite{Dolgov} \\
H(z) & $1.22^{+0.21}_{-0.16}$ & \cite{SKum} \\
SNe Ia & $1.61^{+0.14}_{-0.12}$ & \\

 \hline \hline
\end{tabular}

\caption{\small Best fit values of $\beta$ from other independent studies.}
\end{center}
\end{table}

\begin{figure}[p]
\centering

  \includegraphics[width=1.0\linewidth]{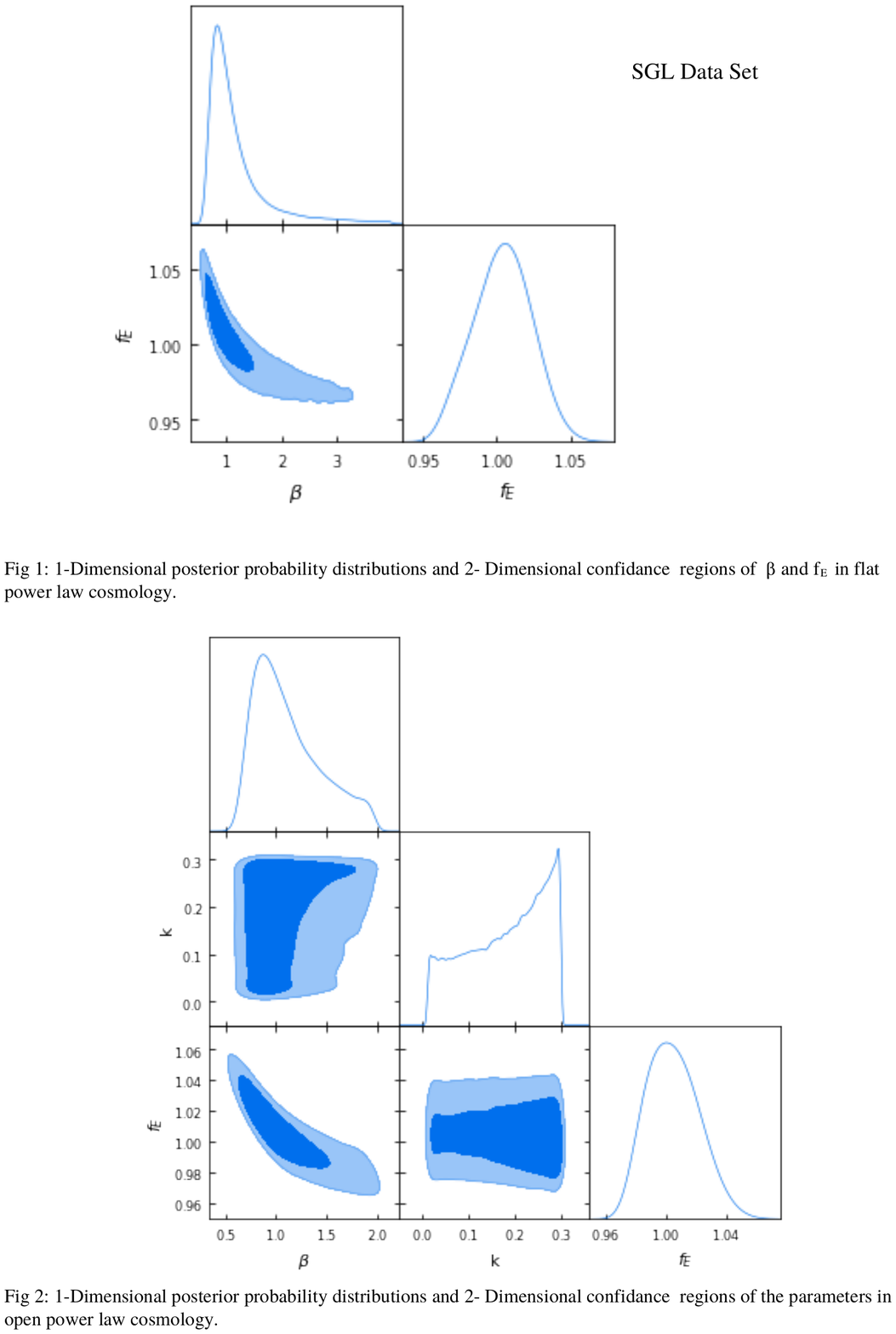}
  
\end{figure}

\begin{figure}
  \includegraphics[width=1.0\linewidth]{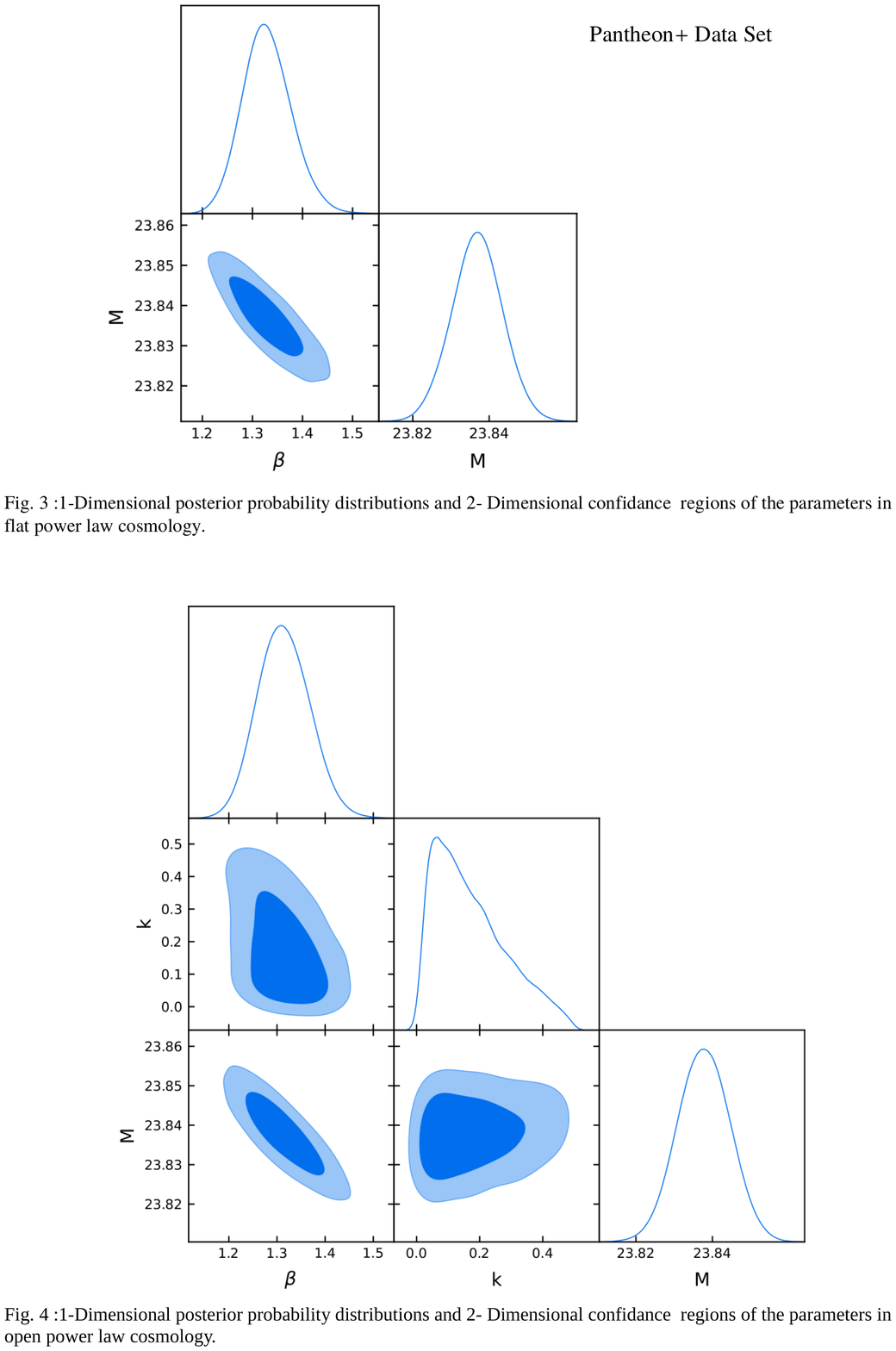}

\end{figure}

\begin{figure}
  \includegraphics[width=1.0\linewidth]{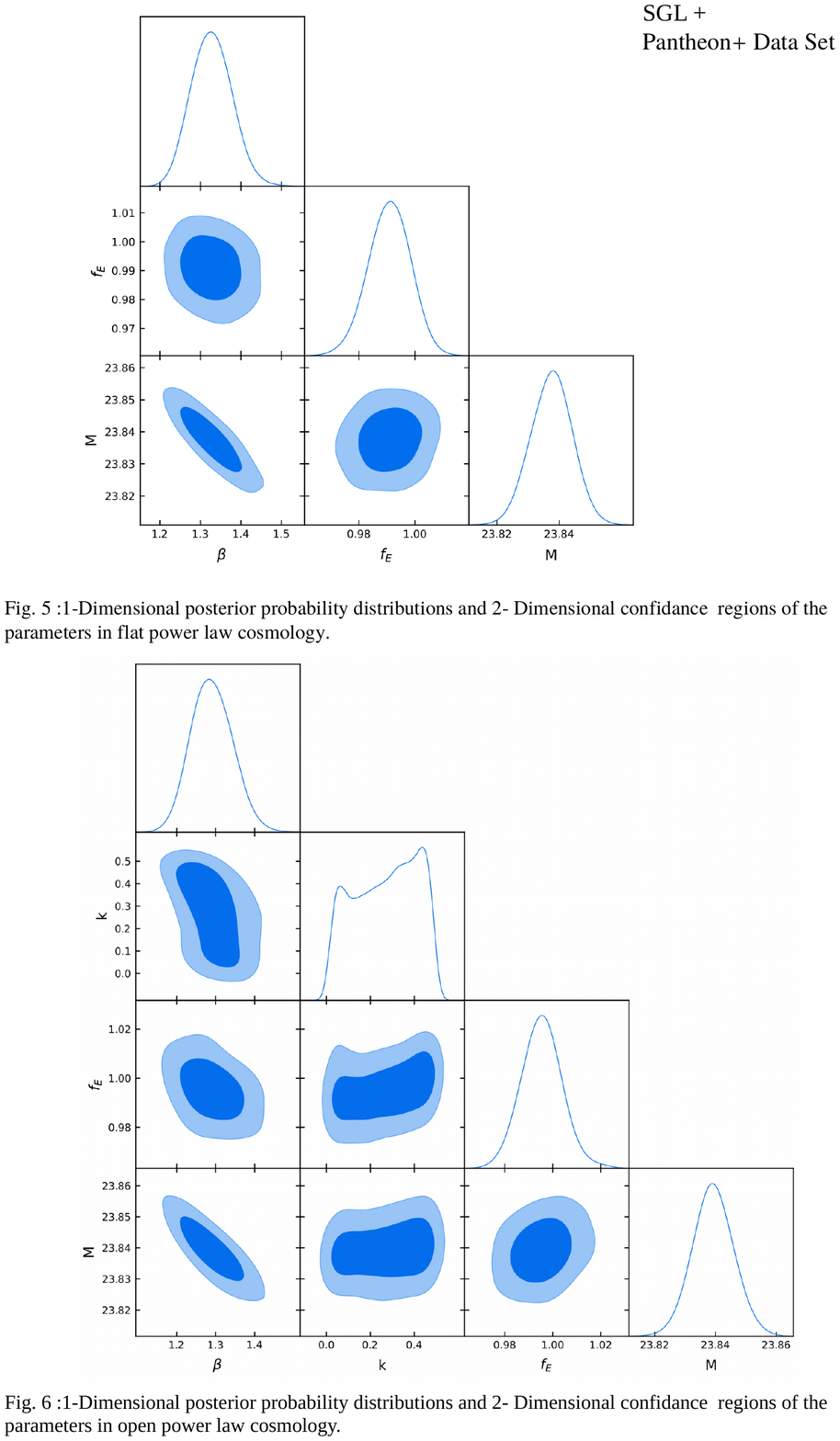}

\end{figure}

\section*{Declarations}

\textbf{Data Availability Statement:} The datasets used in the work are available in public domain. For SGL data, check doi:1509.07649.
Pantheon+ SNe Ia sample is available at:

https://github.com/PantheonPlusSH0ES/PantheonPlusSH0ES.github.io.\\
\textbf{Funding and/or Conflicts of interests/Competing interests} This work is done independently and no funds, grants, or other support was received.  There are no conflicts of interests that are relevant to the content of this article.

\pagebreak

\end{document}